# Study on impact mechanism and precursor information induced by high intensity mining

**Kaiwen Shi[1], Wenhao Shi[1,*], Shankun Zhao[2], Hongfei Duan[3], Yuwei Li[4], Haojie Xue[5], Xueyi Shang[1], Wengang Dang[3], Peng Li[6], Yunfei Zhang[6], Binghuo Guan[6], Xiang Ma[1], Hongke Gao[7]**

1 School of Resources and Safety Engineering, Chongqing University, Chongqing 400044, China
2 China Coal Research Institute, Beijing 100013, China
3 School of Civil Engineering, Sun Yat-sen University, Zhuhai 519082, China
4 School of Environment, Liaoning University, Shenyang 110036, China
5 Shandong University, Jinan 250061, China
6 Shendong Coal Group Corporation, Yulin 719315, China
7 School of Mechanics and Civil Engineering, China University of Mining and Technology-Beijing, Beijing 100083, China
* Correspondence: 17705993236@163.com

**Abstract:** With heightened mining intensity, the incidence of coal bursts is escalating, necessitating advanced understanding and prediction techniques. This research delves into the intricacies of coal burst mechanisms, proposing a novel theoretical model for the release of coal mass energy founded on the tenets of stress superposition. A significant revelation is that the energy culminating in a coal burst is an amalgamation of intrinsic coal strain energy and perturbations from mining activities. Field investigations scrutinize the microseismic parameters across a spectrum of mining velocities, discerning potential failure regions and precursor hallmarks in high-intensity mining environments. Notably, microseismic energy, in such contexts, experiences an augmentation of approximately 2000 J. Numerical simulations executed via 3DEC elucidate stress distribution patterns and failure modalities of adjacent rock structures in relation to mining velocities. The simulations underscore that an uptick in mining speed diminishes the buffer to high-pressure abutments, intensifying inherent pressures. For mitigation, it's advocated that high-intensity mining advances be capped at 11 m/d. Merging theoretical analysis, experimental data, field assessments, and computational simulations, this study proffers a holistic insight into coal burst dynamics, underscoring its value in refining monitoring and early warning protocols in the domain.

## 1. Introduction

Coal burst is a major dynamic disaster faced by coal miners. It is caused by the sudden and violent release of elastic strain energy accumulated from the coal and rock mass [1-3]. Mining depth and mining disturbance intensity on the working face are directly related to coal burst occurrence in high stress concentration zone. In high-intensity mining operations with large-height, rapid mining, and long-wall working faces, the strong mining pressure behavior of the coal mass is more complicated and severe, making it easy to induce dynamic disaster accidents such as coal bursts [4]. Therefore, understanding its triggering mechanism and the stress distribution characteristics of the surrounding rock in high-intensity mining and prevention are very challenging tasks for the improvement of mine safety and productivity.

Over the past few decades, scholars have proposed various hypotheses regarding the mechanism of coal burst. From the perspectives of strength [5], stiffness [6], energy [7, 8], and stability [9], research has been conducted to improve the understanding of the mechanism of coal bursts among coal miners. After extensive engineering practice, more and more coal burst cases have been reported. The researchers understanding of the phenomenon and mechanism of coal bursts are more pertinence (such as fault, coal pillar and hard roof induced coal burst). For instance, He et al. [10] based on the engineering background that strong mine tremors frequently occurred in the 7301-working face of the Baodian coal mine due to thick-hard stratum fracture, a novel method is used to successfully simulate the whole process of tremor generation caused by thick-hard roof fracturing and triggered rockburst in a roadway， the research results provide a useful reference for understanding the mechanisms of rockburst induced by mine tremors. Deng [11] an analytical method was proposed to investigate the effects of various factors on the mechanism of pillar rockburst, such as the amplitude of static and dynamic component loads, frequency, slope, duration, multimode, peakedness, and time gap between two dynamic loads. Li al. [12] based on the elastic–plastic mechanics and damage mechanics, we established the elastic–plastic-brittle mutation rockburst model of coal rock with structural surface, the significance of the study lies in both theory and practice for deeply understanding the rockburst mechanism, as well as for accurate monitoring and early warning. Wang al. [13] one pillar model under different loading stiffness is simulated to assess indicators of pillar burst and the resulting damages, the results show the rockmass damage under soft loading stiffness has larger magnitude of plastic strain work and released energy than that which is under stiff loading stiffness. Jiang [14] and Wang [15] studied the precursor information of fault slip and demonstrated the mechanism of coal bursts induced by fault activation by stimulation of the coal mining process. Although these works analyze the mechanisms of coal burst from different aspects, most of them focus on the process of coal burst under universal mining or deep mining.

The above studies have mainly focused on the impact of faults, coal pillars, and hard roofs on coal bursts. These research subjects may be reactivated due to mining disturbances, thereby inducing coal burst disasters. Previous studies have shown that the conditions leading to coal burst are not only related to coal–rock mass properties (such as strength, stiffness, stability, and energy dissipation) but are also closely related to the mining intensity (Fig. 1 (a) and (b) are the distribution of high-intensity mining mines) [16, 17]. Field investigations indicate that the timbering of surrounding rock roof appears deformed after a week under the influence of high-intensity mining disturbances. Mining intensity usually depends on mining parameters. Such as working face advancing speed, mining height, working face size and mining technology. The main features are sinking, falling, and severe bottom drum phenomenon (Fig. 1, (c) and (d)). The high-intensity mining main phenomena are as follows. 1) A large area of the roadway floor rises and has a visible large-scale fracture. 2) The two sides of the roadway are extruded, and some anchor plates break under the pressure of surrounding rock, causing anchor plate shedding and bolt support failure. 3) Severe sinking or falling of coal seam roof occurs; sinking is approximately 50 cm, and the roof falls under mining disturbances. Owing to the randomness and complexity of conditions leading to coal burst, the coal burst mechanism in areas disturbed by high-intensity mining is unclear. In view of this, gaining insight into the mechanism of coal burst induced by high-intensity mining is necessary.

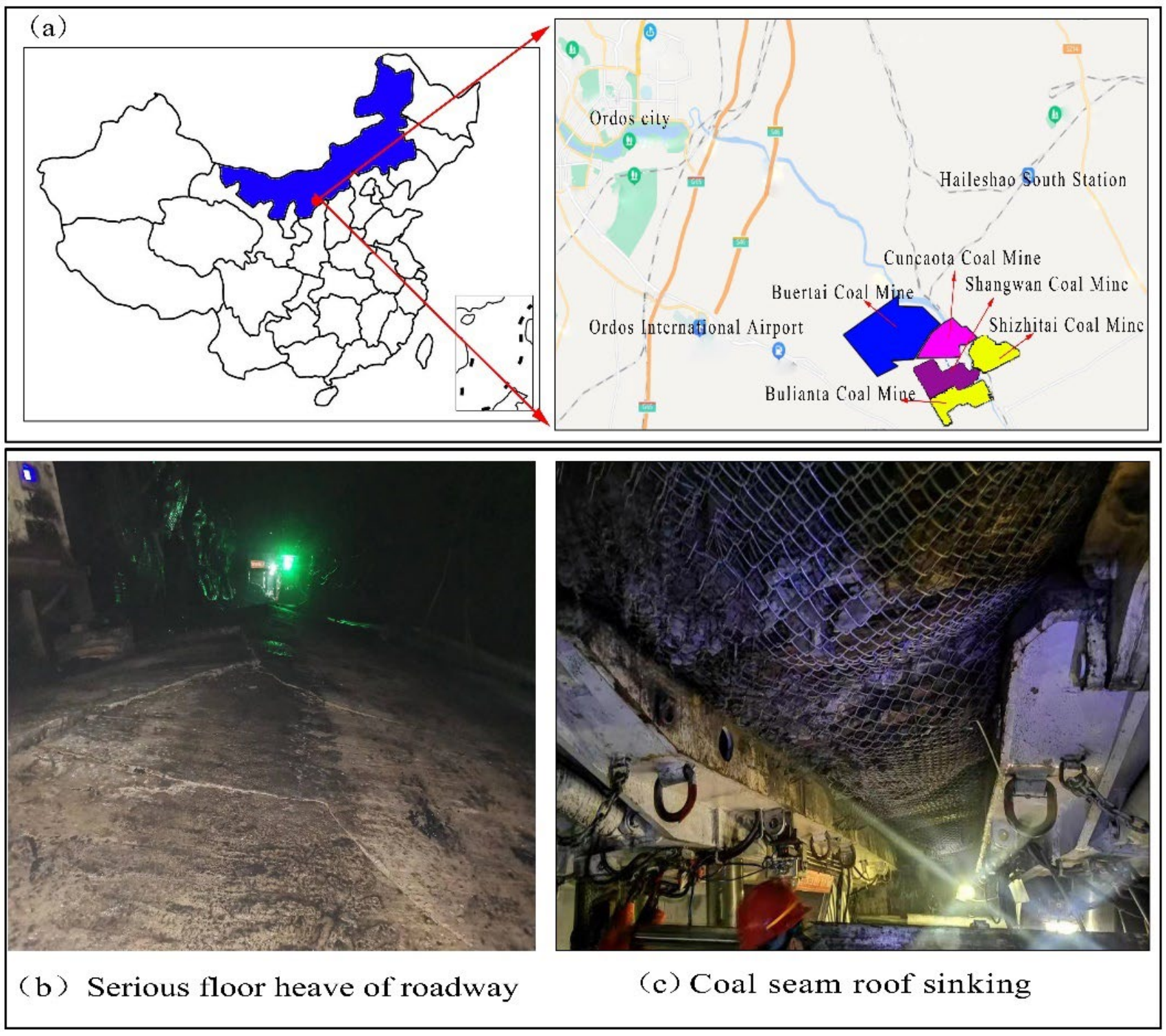

**Figure 1.** Mining area distribution and high-intensity mining pressure behavior diagram

To date, no theory explains coal bursts induced by high-intensity mining in coal seams; this is because of the complexity of coal burst and the diversity of influencing factors. Moreover, the mechanism of coal burst induced by high-intensity mining disturbances remains unclear. No reasonable interpretation explaining the sufficient condition triggering the energy release from surrounding rock induced by high-intensity mining has been presented. Hence, further exploration is necessary to reveal the mechanism of coal burst generated by high-intensity mining accurately.

This study aims to resolve the problems. In this study, the characteristics of microseismic activity parameters under various mining speeds are analyzed, potential failure zones and precursor characteristics are identified during the high-intensity mining. Additionally, the stress distribution characteristics and failure law of the roadway surrounding induced by high-intensity mining are simulated. The results have guiding significance for deeply understanding the mechanism, early warning and prevention of rock burst induced by high-intensity mining.

## 2. Mechanical behavior of surrounding rock under high-intensity mining

Extreme stress due to high-intensity mining can damage part of the surrounding rock mass during coal extraction; the high-intensity mining overburden stress and failure law diagrams are shown in Fig. 2. The interaction associated with this in high-intensity mining fundamentally controls the crack initiation, propagation, and penetration of surrounding rock until macro-fracture occurs, suggesting a link to the mechanism of coal burst induced by high-intensity mining. As shown in Fig. 4, A, B, and C are the stress–strain curves corresponding to the conventional, critical, and high-intensity excavation speeds of the coal mine working face, respectively. When the coal mining rate exceeds the critical excavation speed, the surrounding rock shows considerable plastic deformation, and the coal roof may fracture, exhibiting stochastic characteristics and inducing pressure bumping. If roof fracture occurs in the frontage of the working face and mining continues, roof cutting fracture may occur. However, if the roof fracture occurs behind the working face, then coal mining is safe.

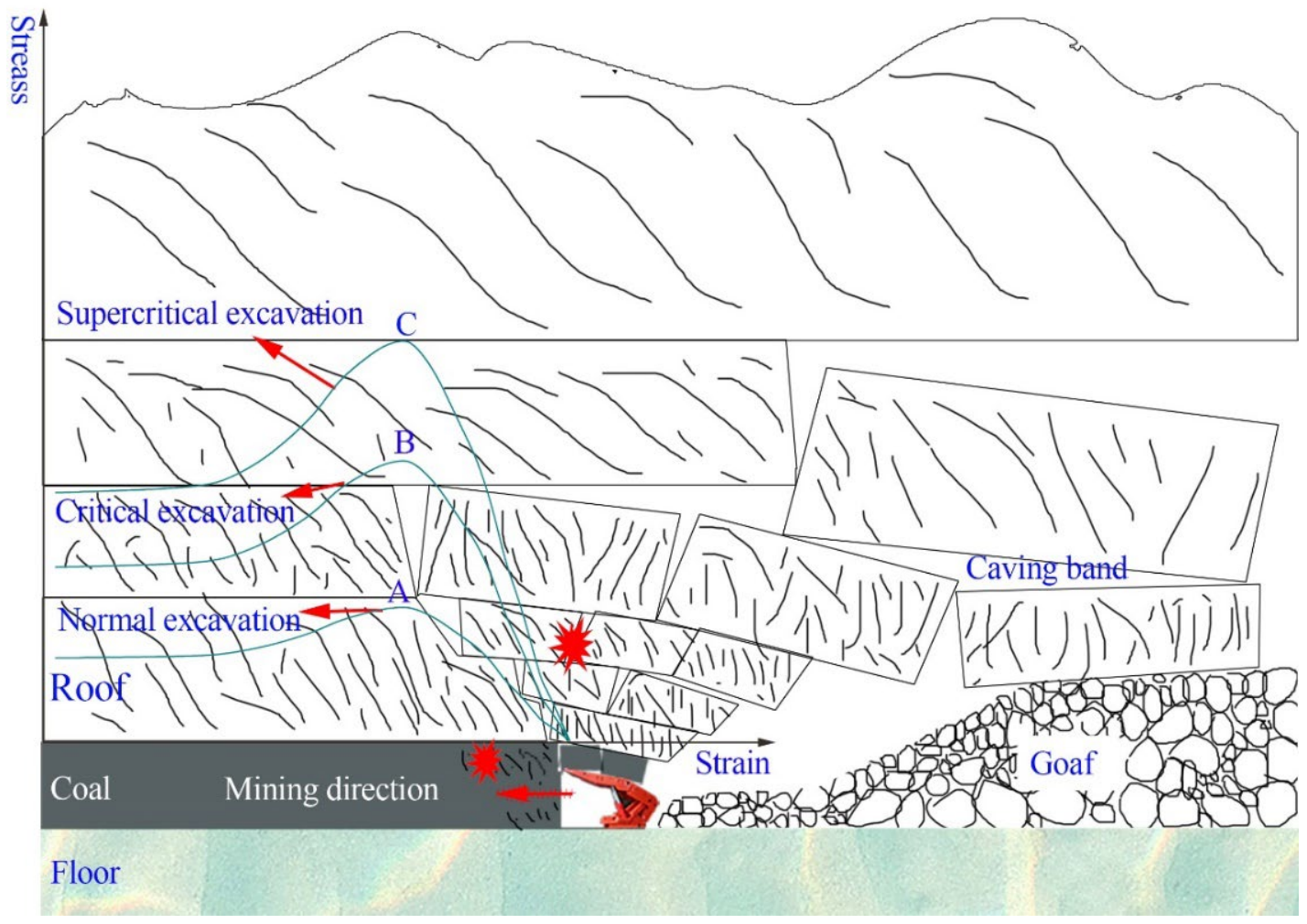

**Figure 2**. High-intensity mining overburden stress and failure law diagram.

During high-intensity mining, numerous microcracks form and penetrate the overlying strata, thereby reducing the rock mass strength. Furthermore, pressure bumping can occur when the superimposed total stress (i.e., initial and high-intensity mining stresses) exceeds the critical strength, $\sigma_c$, of surrounding rock. Accordingly, considering the overlying strata as the research object, one element is used to establish the mechanical analysis model shown in Fig. 3. The stress at the weak interface subjected to high-intensity mining is expressed as follows:

$$\sigma_\alpha = \frac{\sigma_x + \sigma_y}{2} + \frac{\sigma_x - \sigma_y}{2} \cos 2\alpha - \tau_{xy} \sin 2\alpha \tag{1}$$

$$\tau_\alpha = \frac{\sigma_x - \sigma_y}{2} \sin 2\alpha + \tau_{xy} \cos 2\alpha \tag{2}$$

where $\sigma_x$ and $\sigma_y$ are the maximum and minimum principal stresses, respectively; $\tau_{xy}$ is the shear stress in the y direction; and α is the weak interface angle during high-intensity mining. According to the Moore–Coulomb criterion [26], the shear strength, $\tau_f$, of the weak interface is as follows:

$$\tau_f = c - \sigma_\alpha \tan \varphi \tag{3}$$

where $c$ and $\varphi$ denote the cohesion and internal friction angle of rock. By hypothesizing that $\tau_\alpha = \tau_f$, the criteria for assessing the fracture of overlying strata are as follows:

$$\sigma_x - \sigma_y = \frac{2c - \tan\varphi(\sigma_x + \sigma_y) + 2\tau_{xy}(\tan\varphi \sin 2\alpha - \cos 2\alpha)}{\sin 2\alpha(1 + \tan\varphi \cos 2\alpha)} \tag{4}$$

It can be inferred that when $\sigma_x - \sigma_y \to \infty$, the following can be deduced: $\varphi < \alpha < 90°$. When $\alpha = 45° + \frac{\varphi}{2}$, the minimum critical value of the overlying strata fracture is obtained. Equation (4) indicates that the fracture of the overlying strata is related not only to cohesion and internal friction angle but also to the maximum and minimum principal stresses and microcrack initiation angle. In summary, high-intensity mining can readily induce the initiation and propagation of microcracks to the extent of penetrating the overlying strata. When mining is highly intense, coal burst may occur.

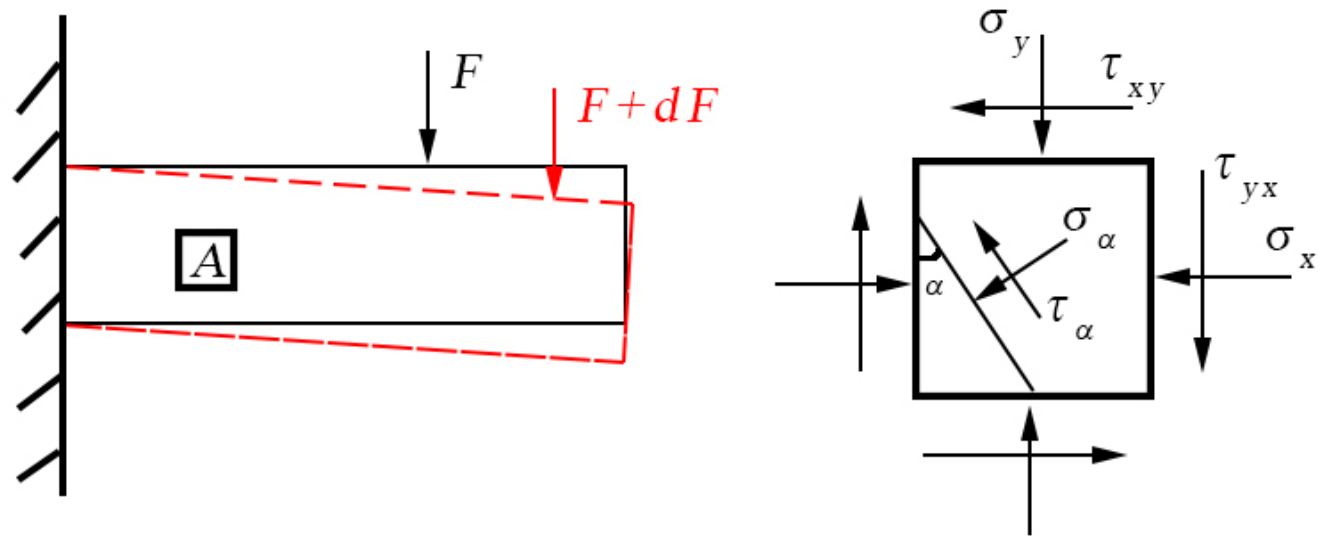

(a) Roof strata adoption mechanical model (b) Stress state diagram

**Figure 3.** Overlying strata fracture mechanics model of high-intensity mining

## 3. Field MS investigation and analysis

### *3.1 Engineering background*

The coal mine is a typical mining area with a large but shallow coal seam with high strength. The mine is situated in Ordos City with the Baotou–Shenmu–Xi'an double-track electrified railway in the east and 210 National Highway in the west. Ordos airport is located approximately 20 km northwest of the mining area. The distribution of mining areas is shown in Fig. 1 (a). The production capacity of the coal seam has exceeded 20 million t/year. The thickness of minable coal seams is 0.90–7.68 m, and the average burial depth is 410 m. The coal seam mining speed reaches 10-15 m/d. In the process of coal seam exploitation, the mine pressure behavior appears violent, which can easily lead to roof puking, spalling, roof cracking, and bottom drum phenomenon (Fig. 1, (b) and (c)).

A comprehensive illustration of column coal and rock layers is presented in Fig. 4. The figure indicates that the roof of working face 42# is sandy mudstone with an average thickness of 14.5 m. The immediate roof is composed of fine mudstone with a thickness of 24.36 m. The floor of 22# coal seam is sandy mudstone with an average thickness of 25 m. The vertical distance between coal seams 42# and 22# is approximately 80 m. The former coal seam is identified to have a weak pressure bumping tendency that is mainly related to the roof and floor lithology. Under high-intensity mining, spalling, roof cracking, and bottom drum phenomenon typically occur.

| number | rock property | property description | shaped | Thickness |
|---|---|---|---|---|
| 1 | siltstone | Gray, hard, mixed composition. | | 9.65 |
| 2 | fine sandstone | White, hard, thick layered, quartz, feldspar - based. | | 35.6 |
| 3 | sandy mudstone | Dark gray, hard, mud-based. | | 11.5 |
| 4 | 42# upper coal | Dark brown, weak asphalt luster, jagged fracture. | | 6.15 |
| 5 | sandy mudstone | Dark gray, hard, mainly argillaceous, followed by sandy. | | 22.5 |
| 6 | 42# coal | With bright coal, dark coal interbedded distribution, semi-bright coal. | | 2.7 |
| 7 | siltstone | Gray white, hard, thick layer, mainly quartz, feldspar. | | 13.2 |

**Figure 4.** Comprehensive column illustration of coal and rock layers in 42# working face.

### *3.2 MS activity characteristics under different mining speeds*

To study the activity characteristics of MS parameters of surrounding rock under different mining speeds. According to the actual exploitation condition of the mining working face, MS signals with mining speeds of 7 m/d, 9 m/d, 11 m/d, 13 m/d, 15 m/d, and 17 m/d were selected for analysis.

Fig. 5 is the evolution characteristics of MS events in the mining working face under different mining speeds. The coal mine is operated work in three shifts. From 8:00 to 0:00, the working face is exploited, and from 0:00 to 8:00, mechanical inspection and roadway Mechanical maintenance are mainly conducted. The number–time curve of MS events with a propulsion speed of 9 m/d is shown in Fig. 5. In Fig. 5(a), the MS events from 8:00 to 0:00 are more frequent than those from 0:00 to 8:00. This is because between 8:00 and 0:00, coal excavation is mainly performed. Coal mining disturbs the surrounding rock; hence, the number of MS events during this period significantly exceeds that in the night shift. Fig. 5 (b) is the characteristic curve of MS events under different mining speed. With the increase of the mining speeds, the MS events increase. When the mining speed exceeds 15 m/d, the number of MS events on the day exceeds 5000, and the number of MS events increases significantly. In general, the number of MS events increases with the increase of mining speed. Meanwhile, the high-energy events under high-speed mining account for the majority, indicating that the mining speed contributes significantly towards the surrounding rock of the roadway. Nevertheless, at 20m in front of the working face, due to the considerable influence of excavation speed on the rock mass, the stress is redistributed and transferred, which leads to the initiation and expansion of microcracks in the rock mass, thus increasing the MS events. Compared with previous research results, MS events under high-intensity mining increase rapidly and are denser. It is also the justification why the periodic pressure is severe, and the bottom drum and the rib spalling are serious.

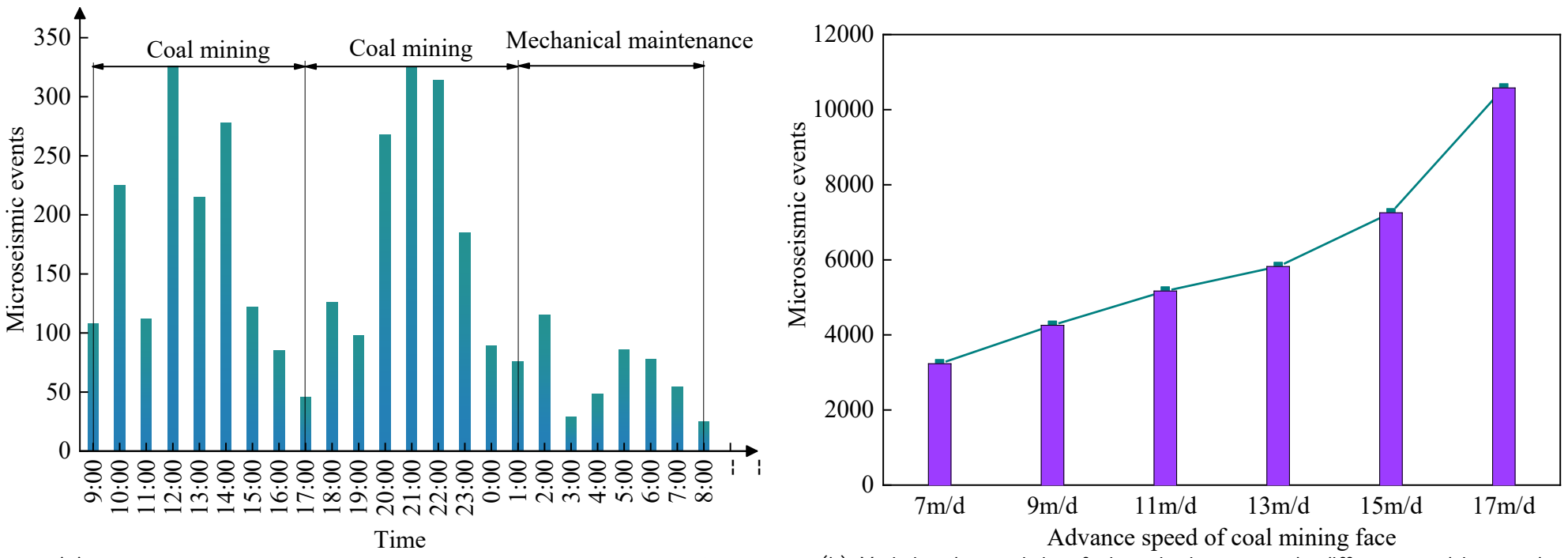

(a) Microseismic event-time distribution with a velocity of 9m /d.

(b) Variation characteristics of microseismic events under different propulsion speeds

**Figure 5.** Evolution characteristics of microseismic events at different advancing speeds in coal mining face.

Fig.6 is the energy-time curve of coal mining face under different advancing speeds. The figure indicates that high-intensity mining induces rock mass to release massive amounts of energy, and the stability of surrounding rock is inadequate. With the increase of coal mining speed, MS high-energy events are increasing. It indicates that the MS high-energy events under high-intensity mining are more frequent than those under general mining speed and more intensive. Figure 9 shows in the energy variation characteristics under different mining speeds. By analyzing the energy characteristics of diverse sections (Fig. 7 (a)), it is found that the proportion of microseismic small energy occurrence is high under low mining speeds. It can be observed in Fig. 7 (b) that the frequency of microseismic high-energy events increases with the increase of mining speed. It shows that the mining speed can trigger more microseismic high-energy events and increase the risk of coal burst. Under the influence of high-intensity mining, the instability and failure of rock mass, the initiation and propagation of microcracks, and the energy dissipation of rock mass increase the risk of coal burst. The prediction results indicate that the elastic strain energy released by coal–rock mass increases under high-intensity mining, and the MS energy suddenly increases during coal mining. Compared with previous research results, the MS energy under high-intensity mining is higher by approximately 20 00 J. This reveals that the microcrack initiation and propagation under high-intensity mining are severe, and the stability of the surrounding rock is inadequate.

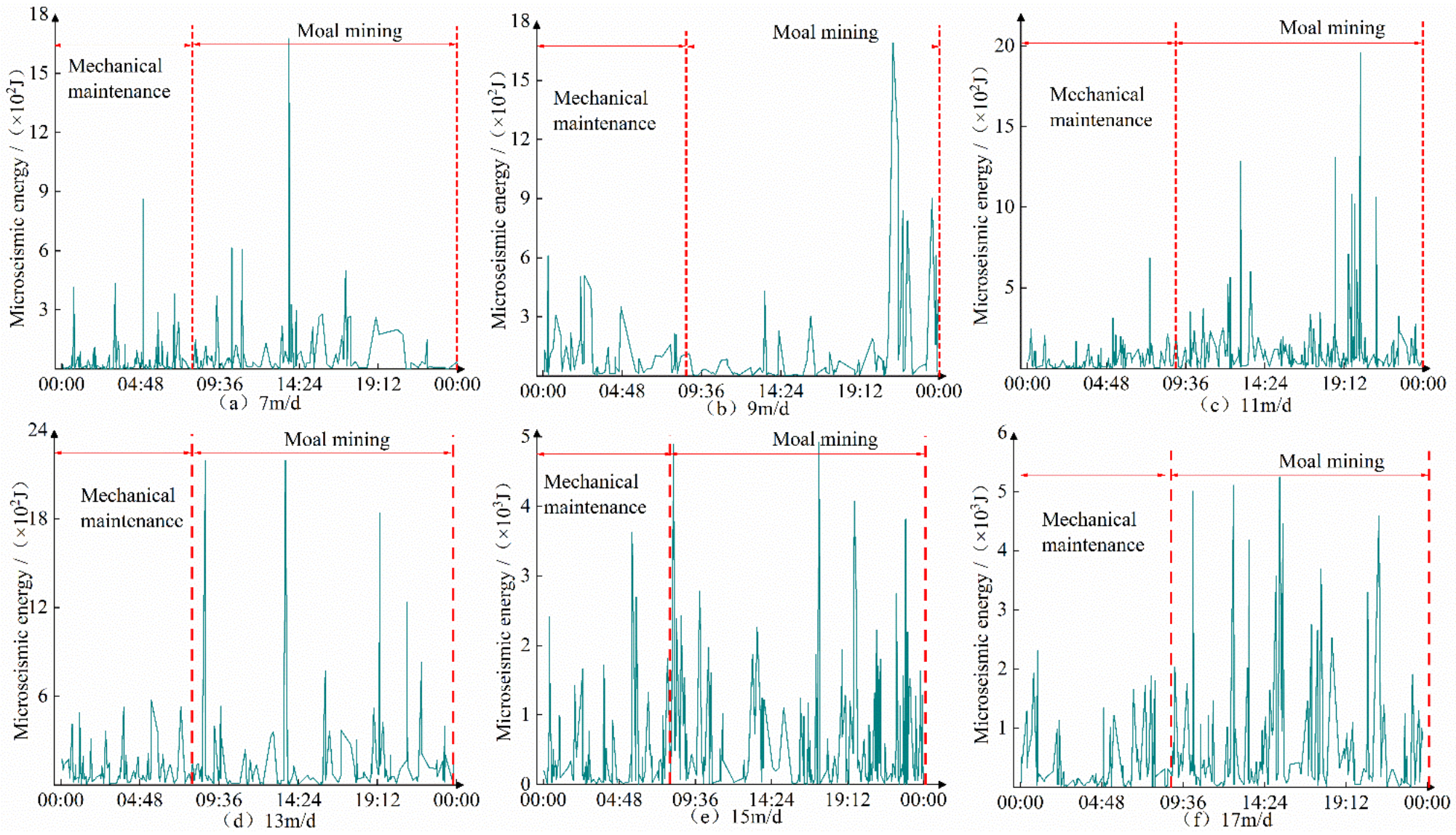

**Figure 6.** The distribution curve of MS energy with time when the advancing speed of coal face. a is 7m / s, b is 9m / s, c is 11m / s, d is 13m / s, e is 15m / s, f is 17m / s.

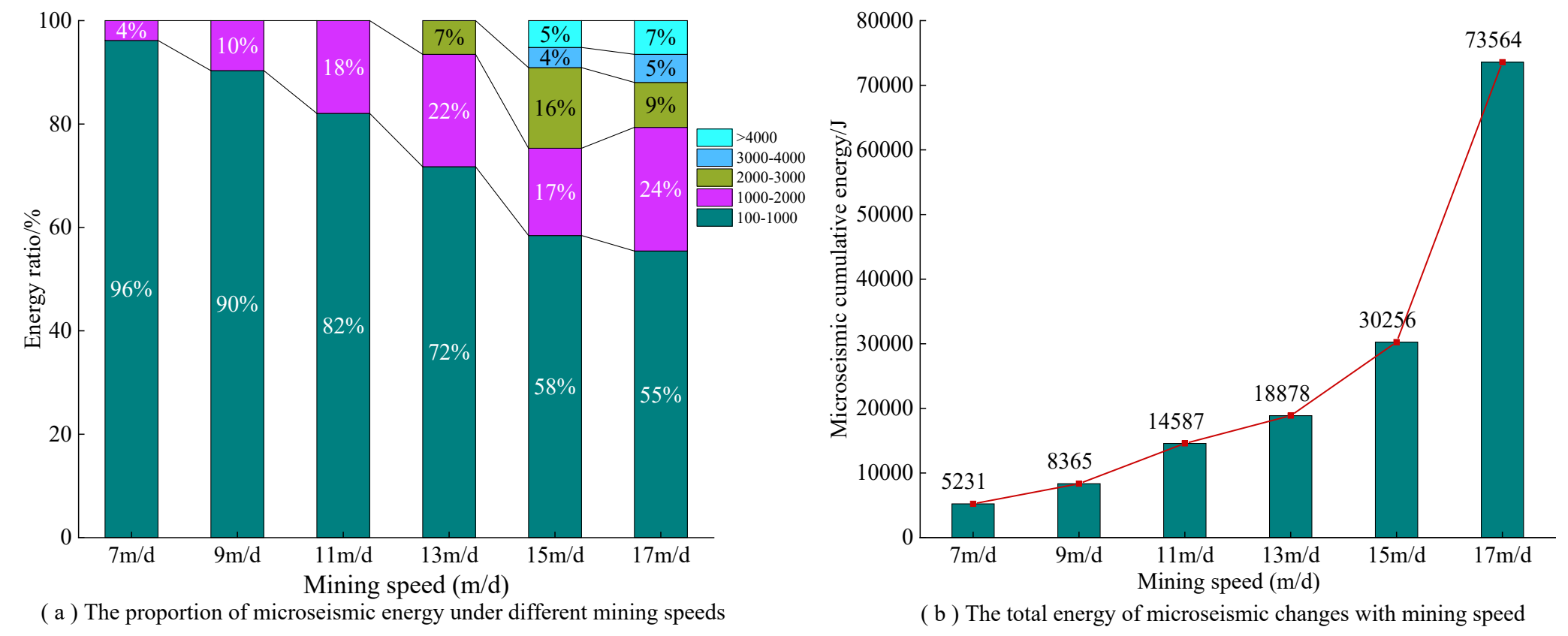

**Figure 7.** Characteristics of energy changing with mining speed

The MS energy index is defined as the ratio of the observed radiated MS energy to the average radiated energy of the same seismic moment of the event, which is defined as follows:

$$EI = \frac{E}{\overline{E(M)}} \tag{5}$$

where $EI$ is energy index, $E$ is the energy of a MS event, $\overline{E(M)}$ is the average radiation energy of seismic moment $M$, The relationship between the average MS energy $\overline{E(M)}$ and the MS energy $E(M)$ in the study area is as follows:

$$\lg \overline{E(M)} = c + d \lg M \tag{6}$$

where c and d are constants. It is based on the relationship between the energy released by the microseismic and the seismic moment, as shown in Figure 8. The MS energy index can reflect the stress distribution in the monitoring area. The increase of energy index means that the stress of surrounding rock increases, indicating that the risk of rock instability increases.

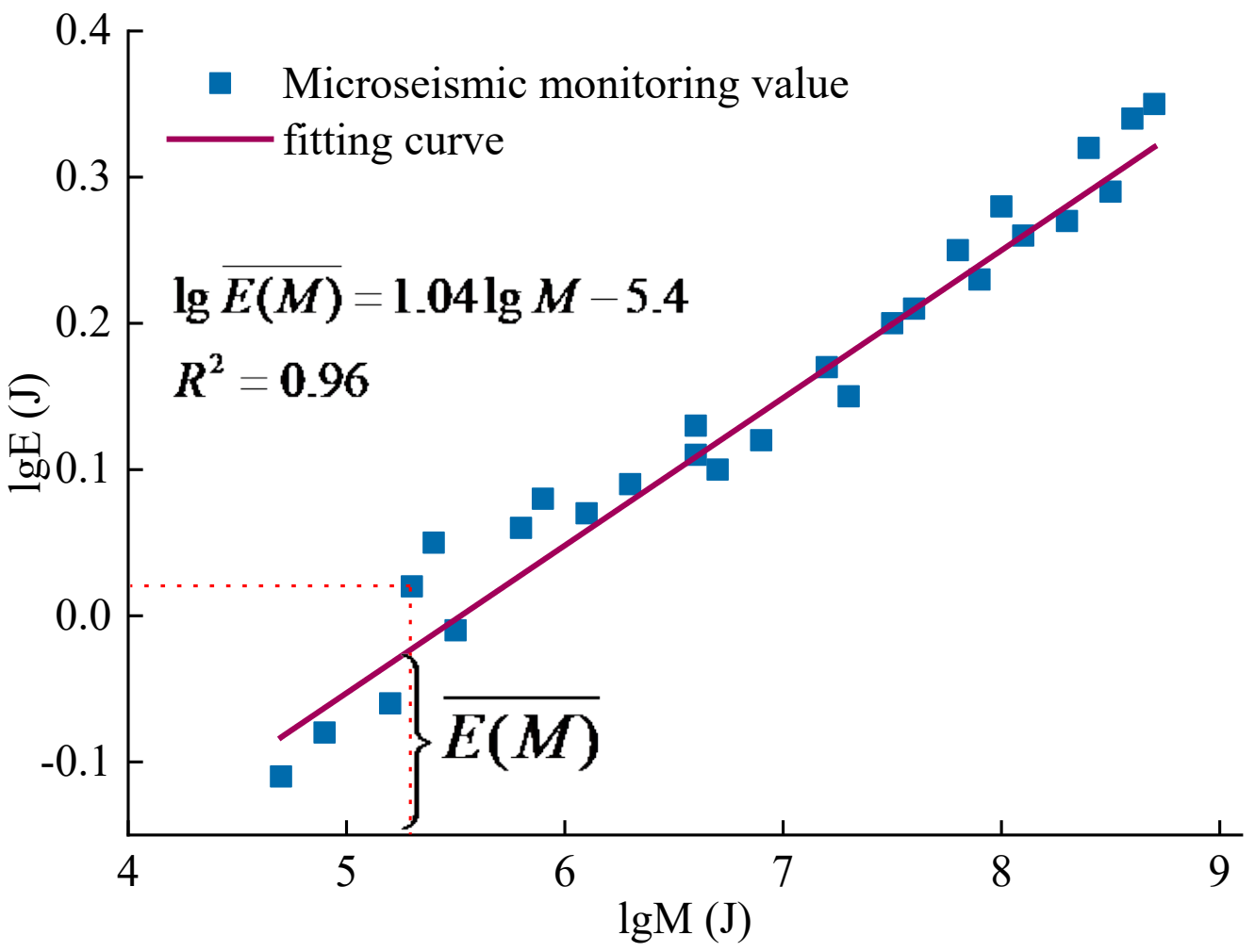

**Figure 8.** Relationship between microseismic energy and seismic moment

The inelastic strain of rock mass is expressed by apparent volume, and the expression is as follows:

$$V_A = \frac{M}{2\sigma_A} \tag{7}$$

where $V_A$ is the volume of rock mass, $\sigma_A$ is the apparent stress, $M$ represents seismic moment.

Fig. 9 is curves of cumulative apparent volume, energy index, and cumulative release energy of microseismic events at different mining speed. The cumulative apparent volume and cumulative release energy increase slowly in the small period of mining speed, without mutation, and the fluctuation range of energy index curve is small. When the mining speed exceeds 15 m/d, the energy index curve rises suddenly and rapidly, while the cumulative apparent volume increases slowly, indicating that the surrounding rock is in the stage of energy accumulation and release. The result indicates that under high-intensity mining, the energy index curve rises sharply again, the cumulative apparent volume gradually increases, and the local surrounding rock stress transfers to the nearby surrounding rock, forming a new stress concentration area.

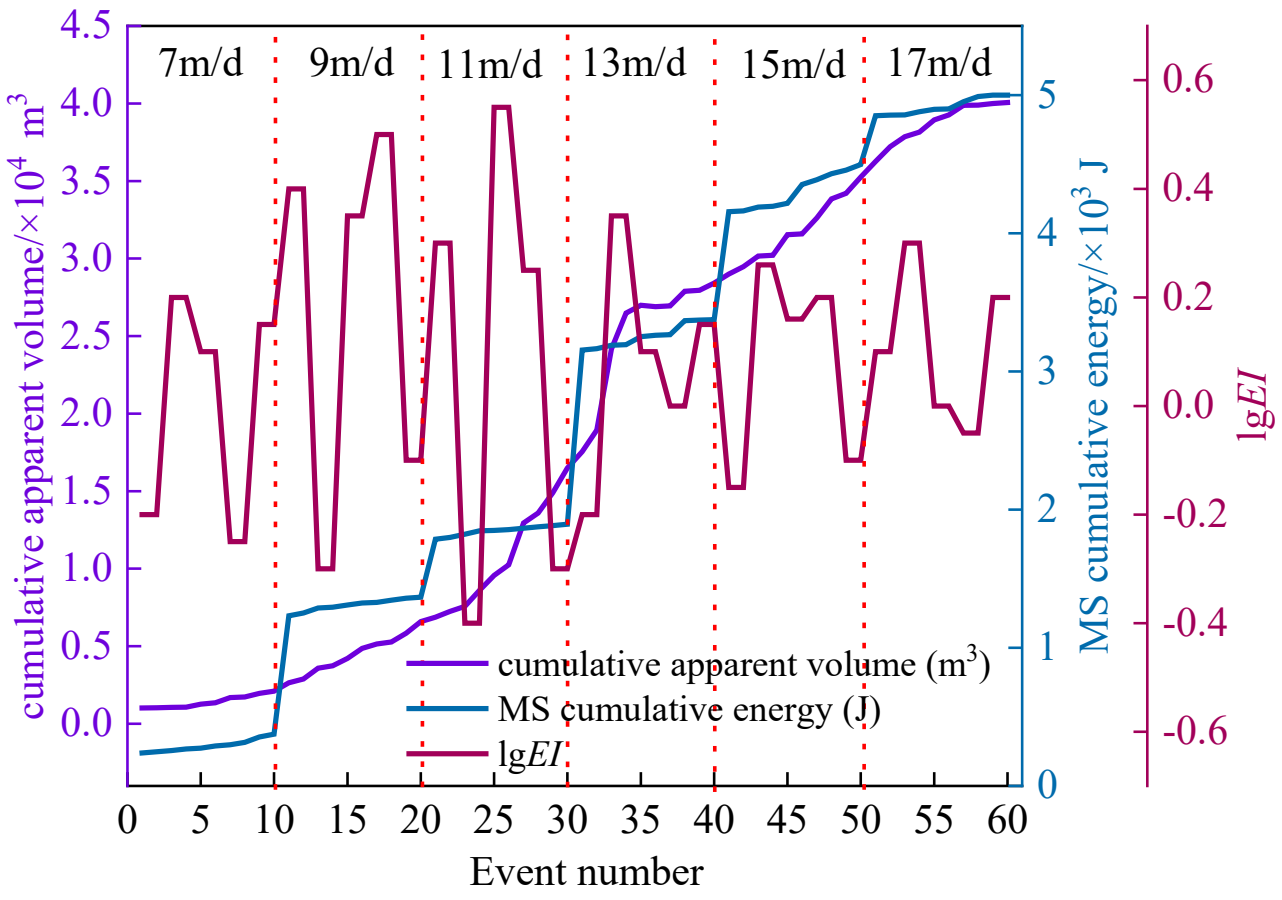

**Figure 9.** curves of cumulative apparent volume, energy index, and cumulative release energy of microseismic events at different mining speed

In summary, the increase in mining speed leads to the increase in coal burst risk. In general, the damage and elastic strain energy in the coal mass accumulate continuously as the advancing speed increases, providing energy conditions for the occurrence of coal burst. Therefore, under high-intensity mining, the greater the risk of coal burst is, which is consistent with the laboratory test results in the second section. In addition, when the mining rate exceeds 11m/d, brittle failure and roof spalling occur on the advance roadway. Accordingly, to ensure the safe and efficient mining of coal mines, the design mining speed of the working face of the mine should not exceed 11m/d.

## 4 Numerical analysis of different excavation speeds

### *4.1 Rock mechanics parameters and excavation scheme*

Based on the geological condition investigation of 42# working faces, it has carried on the numerical simulation of working face excavation. Different mining speeds of coal mining faces have been simulated by commercial software of discrete element simulation. Fig. 10(a) is the top view of 42204# coal mining face.

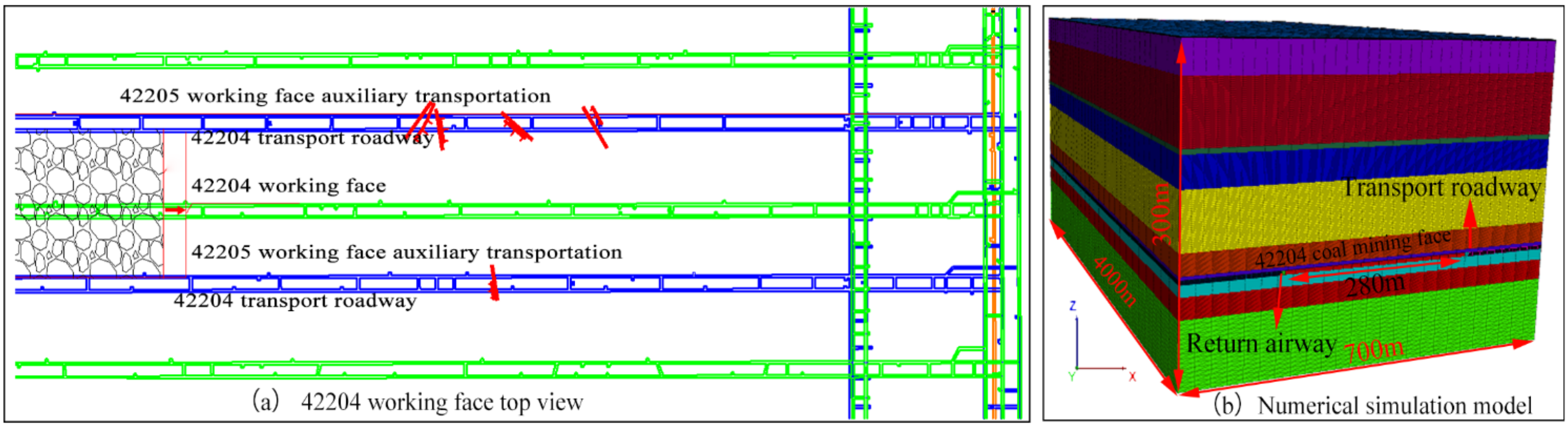

**Figure 10.** 42204# working face top view and numerical simulation model

Table 1 shows the mechanical parameters of simulated rock strata. This paper considers Six kinds of lithologic strata in the simulation. The model size is designed to be 4000m × 700m × 300m. Gravity is applied above the model, displacement is limited in X and Y directions, and the bottom edge is fixed. The numerical model is shown in Fig.12(b). The simulated mining speed is 7 m/d, 9 m/d, 11 m/d, 13 m/d, 15 m/d, 17 m/d, respectively.

**Table 1.** Rock mechanics parameters

| Lithology | Thickness (m) | Poisson ratio | Elastic modulus （GPa） | Tensile strength （MPa） | Cohesion （MPa） | Angle of internal friction （°） |
|---|---|---|---|---|---|---|
| Medium grained sandstone | 46 | 0.36 | 1.5 | 1.58 | 1.6 | 29 |
| Gritstone | 50 | 0.45 | 2.1 | 1.5 | 1.17 | 27 |
| Siltstone | 12 | 0.38 | 1.8 | 1.52 | 1.9 | 28 |
| Gravel coarse sandstone | 44 | 0.64 | 3.3 | 5.6 | 2.82 | 27 |
| Mudstone | 61 | 0.72 | 5.2 | 7.57 | 3.1 | 29 |
| Medium grained sandstone | 30 | 0.82 | 4.5 | 5.2 | 1.4 | 26 |
| Siltstone | 5 | 0.31 | 3.3 | 1.51 | 1.2 | 25 |
| 42# coal seam | 6 | 0.25 | 0.95 | 1.2 | 0.82 | 24 |
| Medium grained sandstone | 16 | 0.35 | 1.9 | 2.1 | 3.1 | 28 |
| Mudstone | 30 | 0.55 | 3.6 | 3.6 | 4.2 | 29 |

### *4.2 Analysis of simulation results*

Fig. 11 is the stress change and deformation characteristics of coal face under different advancing speeds. When the mining speed are 7m/d, 9m/d, 11m/d, 13m/d, 15m/d and 17m/d, the abutment pressure is 29.3MPa, 32.7MPa, 34MPa, 37MPa, 41.2MPa and 47.8MPa respectively. It is demonstrated that the abutment pressure increases with the increase of mining speed. The increasing range were 11.6 %, 3.97 %, 8.82 %, 11.35 % and 16 %, respectively. In addition, with the increase in mining speed, the displacement of overlying strata can be observed from the displacement cloud. It demonstrates that the greater the mining speed of the working face, the greater the disturbance to the stability of the surrounding rock of the roadway. Meanwhile, when the mining speed exceeds 11 m/step, combined with MS monitoring, it is found that the elastic strain energy accumulated in the surrounding rock of the roadway is increasing. Even if the mining disturbance is trivial, it is easy to induce the instantaneous release of the elastic strain energy accumulated in the surrounding rock. With subjected to abutment pressure, the primary crack of the roadway surrounding rock expands and penetrates, which finally leads to the instability and failure of the surrounding rock. Meanwhile, the mining disturbance is very small and it is easy to induce the instantaneous release of the elastic strain energy accumulated in the surrounding rock of the roadway. When the surrounding rock exceeds its own strength, the elastic strain energy is released instantaneously, which leads to the appearance of strong mine pressure under dynamic load, which is also the root cause of coal burst induced by high-intensity mining.

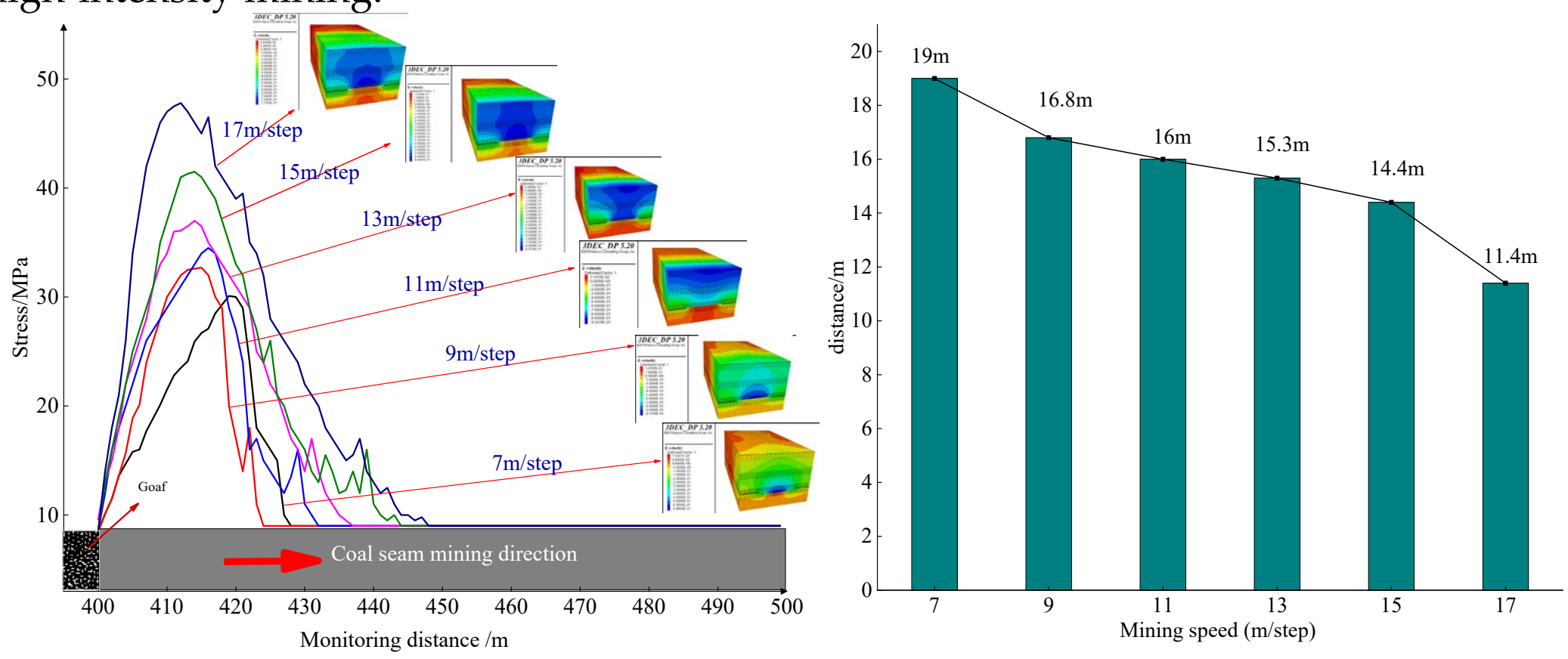

(a)Stress change and deformation characteristics of coal face under different advancing speeds

(b)Abutment pressure distance working face distance

**Figure 11.** Stress change and deformation characteristics of coal face under different advancing speeds

Fig.12 shows the stress distribution characteristics of the roadway surrounding rock when the working face advances 400 m under high-intensity mining. According to Figure 14, it can be observed that with the increase in mining speeds, the mining stress distribution area of the roadway surrounding rock increases. Comparing different mining speeds, when the mining speed reaches 15m/step, the mining-induced stress disturbance length to the roadway reaches 331m, and the disturbance area to the coal seam reaches 227m. Compared with the mining speed of 7m / step, the mining-induced stress disturbance length of the roadway is increased by 93.56 %, and the disturbance of each layer is increased by 157.9 %. When the mining speed is 7m / d, 9m / d, 11m / d, 13m / d, 15m / d and 17m / d, the abutment pressure is 19m, 16.8m, 16m, 15.3m, 14.4m and 11.4m respectively. The above analysis shows that the mining speed has a significant effect on the stability of the surrounding rock of the roadway. The specific performance is that the high-intensity mining has a wide range of disturbance to the surrounding rock of the roadway, and the distance between the super-strong abutment pressure and the working face is reduced. It shows that when the coal mine reaches high-intensity mining, it is easy to induce coal burst accidents of roadway surrounding rock near the working face.

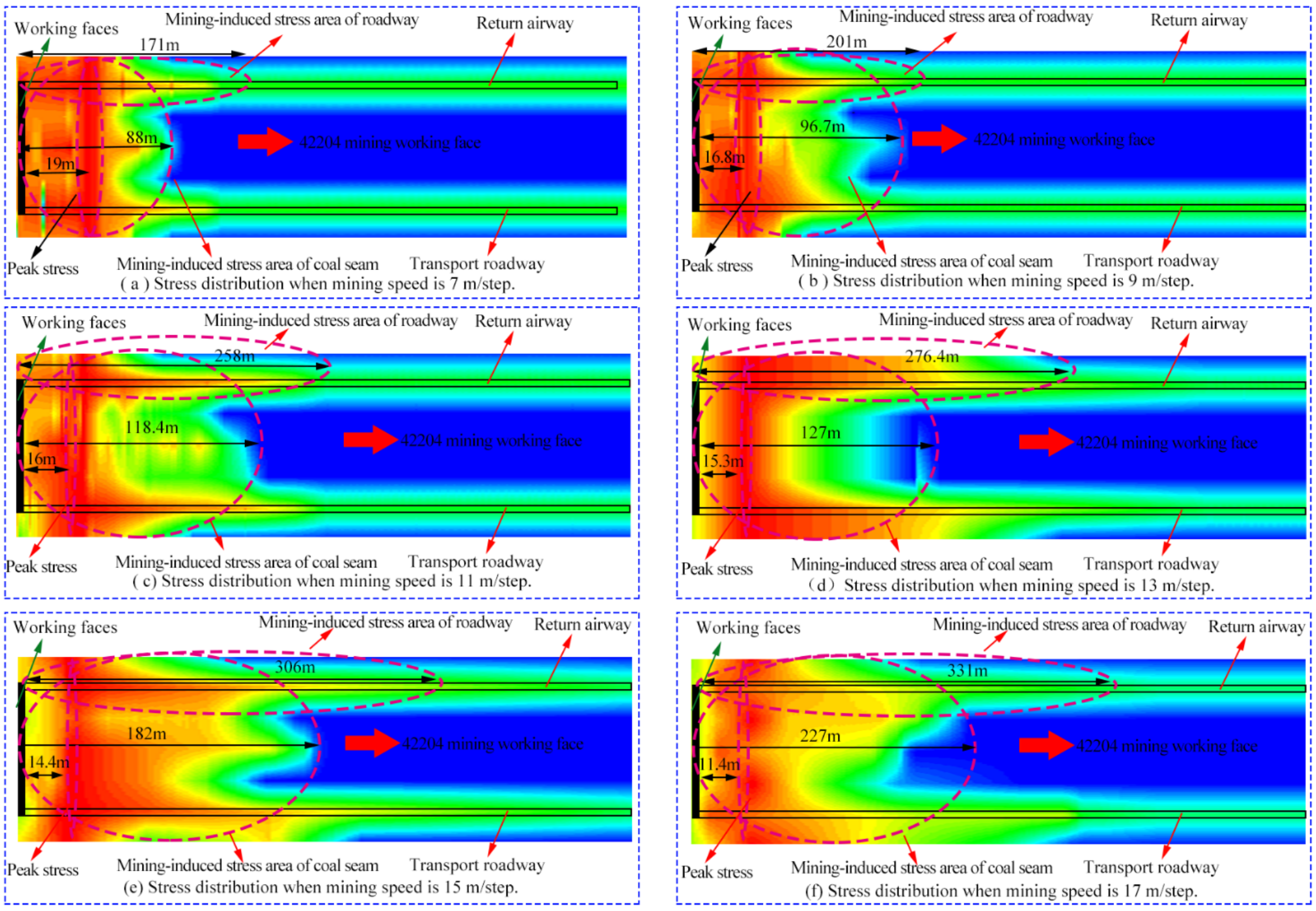

**Figure 12.** The stress distribution characteristics of working face advancing 400 m

In summary, the mining speed has a prominent influence on the stability of the roadway surrounding rock. Meanwhile, it deemed that the increase of mining speed leads to augmentation of stress concentration in front of the working face, and the stress concentration area is given to the brink of the working face with the increase of mining speed. The stress concentration of coal body leads to is to strengthen elastic strain energy. When the elastic strain energy exceeds the bearing limit of coal body, it may induce dynamic disasters such as coal burst. Based on the comprehensive analysis of the geological background of the mine, microseismic monitoring and numerical analysis, the reasonable mining speed is proposed. Considering the relationship between the on-site determination of the propulsion speed and the maximization of safety production benefits, it is finally proposed that the mine 's propulsion speed should be designed to be 11 m/d.

## 4. Discussion

The research shows that the mining overburden rock is subjected to loading and unloading stress for a long time [20, 25, 26]. The instability failure of coal-rock mass induced by high-intensity mining is essentially the crack initiation and propagation of the specimen under cyclic loading and unloading. Consequently, when the sample breaks, a distinct brittle sound can be heard, and a residual heat sensation may be felt. This explanation illustrates that coal bursts can be recognized by the sound of explosions, accompanied by the ejection of debris or rocks. Additionally, the temperature in the affected area typically experiences a noticeable increase.

The stability of the disturbed coal-rock mass decreases as the mining intensity increases. The mining intensity is typically determined by several factors, including the thickness of the coal seam, the speed at which the working face advances, and the length of the working face. Under a specific mining height and working face length, it becomes increasingly evident that the impact of mining on the overlying rock intensifies as the advancing speed accelerates. The energy release during the fracture and instability of coal–rock mass under high-intensity exploitation, known as an MS event, is captured by the MS system [27-29]. Based on the data presented in Figure 13, it is evident that an increase in the advancing speed correlates with a proportional extension of the impact area caused by the tunnel and the coal seam. Combined with Figures 7 and 8, the relationship between the advancing speed of the working face and the intensity of microseismic events becomes apparent. As the advancing speed increases, the microseismic events become more pronounced. This observation illustrates that high-intensity mining disturbance has a significant impact on the stability of the surrounding rock. Under the disturbance of high-intensity mining, the elastic strain energy released by coal-rock mass increases. During the construction of the working face, the microseismic energy is high. Compare to previous research results, the MS energy under high-intensity mining is approximately 20 00 J higher. This indicates that under high-intensity mining, the initiation and propagation of microcracks are particularly severe, exposing the inadequate stability of the surrounding rock. Moreover, this is the primary factor contributing to the intensified occurrence of spalling, roof cracking, and bottom drum phenomenon as the time of periodic pressure shortens and the roof weighting pace slows down.

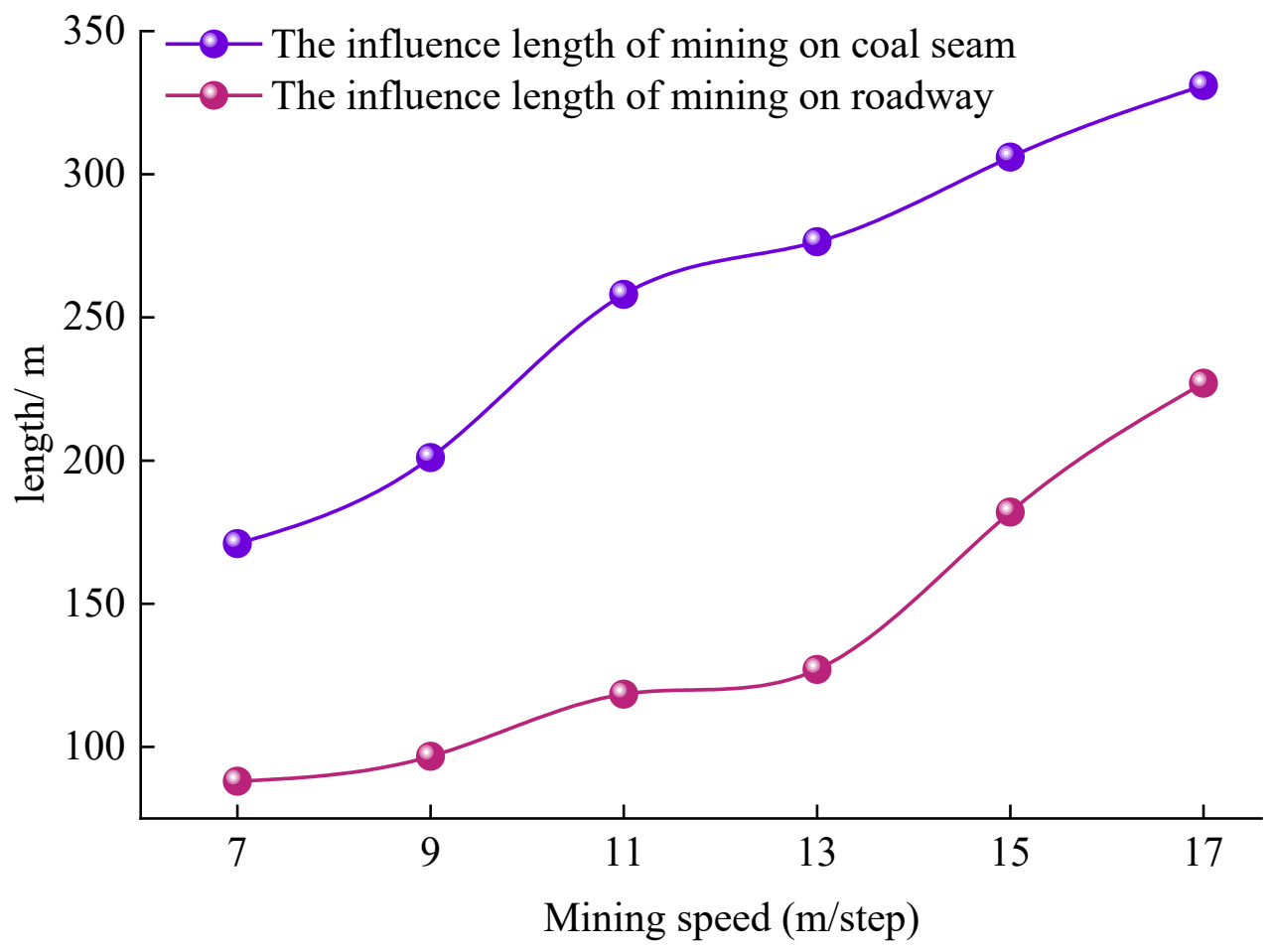

**Figure 1.** Influence of different mining speed on roadway and coal seam

The high-intensity mining operation is the dominant factor that induces the instability of the coal-rock system. Due to the meso-heterogeneity of the coal-rock mass, numerous micro-defects such as micro-cracks and micro-holes are formed during the coal-forming process. Different mining intensities have varying impacts on the disturbance of coal-rock mass. During high-intensity mining, a large area of the overlaying rock becomes unsupported, resulting in the accumulation of elastic strain energy within the coal-rock system. When the coal-rock system reaches its ultimate strength, even minor mining disturbances can trigger a sudden and violent release of the accumulated elastic strain energy within the coal-rock system. Simultaneously, a significant amount of coal-rock mass is being ejected into the goaf, leading to the eventual triggering of instability and the subsequent destruction of the coal-rock mass through impact. The findings suggest that mining intensity plays a pivotal role in impacting ground pressure and serves as the primary causative factor contributing to the instability and destruction of the coal-rock mass.

## 5. Conclusions

Pressure bumping induced by high-intensity mining is fundamentally a fatigue damage accumulation process in coal rock under cyclic loading and unloading. This indicates the dynamic impact instability of the rock mass induced by additional stress superposition. High elastic strain energy is released when the rock mass fractures, and coal outburst propensity is strong when the rock mass rupture is severe.

Under the influence of high-intensity mining, the energy dissipation of rock mass increases the risk of coal burst. Meanwhile, the MS energy under high-intensity mining is higher by approximately 2000 J. In this context, it is suggested that the working face in high-strength mining should advance at a reasonable speed of 11 m/d.

The simulation results show that the high-intensity mining has a wide range of disturbance to the surrounding rock of the roadway, and the distance between the super-strong abutment pressure and the working face is reduced. Proposing the reasonable mining speed under high strength mining by analyzing numerical simulation, On-site microseismic monitoring and laboratory experiment.

**Author Contributions:** Conceptualization, Na Tan and Kaiwen Shi; methodology, Jie CHEN; software, Xiaoying Li and Zhuqing Li; formal analysis, Xiao Wu; resources, Yu Zhang; data curation, Kaiwen Shi; writing—original draft preparation, Kaiwen Shi. All authors have read and agreed to the published version of the manuscript.

**Funding:** This research was funded by National Natural Science Foundation of China, grant number 52474093, the Central Universities Basic Research Funding Projects, grant number 2022CDJXY-009, the Huaneng Group headquarters technology projects, grant number HNKJ22-HF122, the National Natural Science Foundation of China, grant number 52304123, the Postdoctoral Fellowship Program of CPSF, grant number GZB20230914, the China Postdoctoral Science Foundation, grant number 2023M730412.

**Data Availability Statement:** The datasets generated during and/or analysed during the current study are available from the corresponding author on reasonable request.

**Conflicts of Interest:** The authors declare no conflicts of interest.